\documentclass{iau} 
\usepackage{graphicx}

\title[IAUS334.~~Galactic Archeology with RAVE and TGAS] 
{Galactic Archeology with RAVE and TGAS}

\author[Matthias Steinmetz \& the RAVE collaboration]   
{Matthias Steinmetz 
 \and the RAVE collaboration}

\affiliation{Leibniz-Institut f\"ur Astrophysik Potsdam (AIP),\\An der Sternwarte 16, 14482 Potsdam, Germany, email: {\tt msteinmetz@aip.de}}

\pubyear{2017}
\volume{334}  
\jname{Rediscovering our Galaxy}
\editors{C. Chiappini, I. Minchev, E. Starkenburg \& M. Valentini, eds.}
\begin{document}

\maketitle

\begin{abstract}
The 5th RAVE data release  is based on 520,781 spectra ($R\approx7500$ in the CaT region at $8410$ - $8795$\AA) of 457,588 unique stars.  RAVE DR5 provides radial velocities, stellar parameters  and individual abundances for up to seven elements and distances found using isochrones for a considerable subset of these objects. In particular, RAVE DR5 has 255,922 stellar observations that also have parallaxes and proper motions from the Tycho-Gaia astrometric solution (TGAS) in Gaia DR1. The combination of RAVE and TGAS thus provides the currently largest overlap of spectroscopic and space-based astrometric data and thus can serve as a formidable preview of what Gaia is going to deliver in coming data releases. Basic properties of the RAVE+TGAS survey and its derived data products are presented as well as first applications w.r.t wave-like patterns in the disk structure. An outlook to the 6th RAVE data release is given.
\end{abstract}

\firstsection 
\section{Introduction}

Deciphering the structure and formation history of the Galaxy provides important clues to understanding galaxy formation in a broader context.  Wide field spectroscopic surveys play a particularly important role in the analysis of the Milky Way: Spectroscopy enables us to measure the radial velocity, which in turn allows us to study the details of Galactic dynamics. Spectroscopy also permits to measure the abundance of chemical elements in a stellar atmosphere which holds important clues on the initial chemical composition and its subsequent metal enrichment. Despite this importance, ten years ago wide-field spectroscopic surveys of the Milky Way was still limited to the Geneva Copenhagen survey (GCS, Nordstr\"om et al, 2004), which only covered a sphere of about 100~pc radius around the sun (the so-called Hipparcos sphere).

The situation has fundamentally changed over the past decade, with several wide-field spectroscopic surveys underway: SDSS-SEGUE and RAVE being completed in terms of data taking, LAMOST, APOGEE and HERMES well underway, and some massive campaigns such as 4MOST, WEAVE and DESI in the making, each of the latter delivering spectroscopic data for some 10 million stars. The Gaia mission will also not only provide exquisite distances and proper motions for up to 2 billion stars, it also will deliver spectra with resolution and signal-to-noise similar to that of RAVE for some 100 million objects\footnote{For a visualization of the volume covered by some of these surveys, see http://www.rave-survey.org/project/gallery/movies/}. The RAVE survey can play a particular role in the preparation of the Gaia era.

\begin{figure}[t]
\begin{center}
 \includegraphics[width=4.5in]{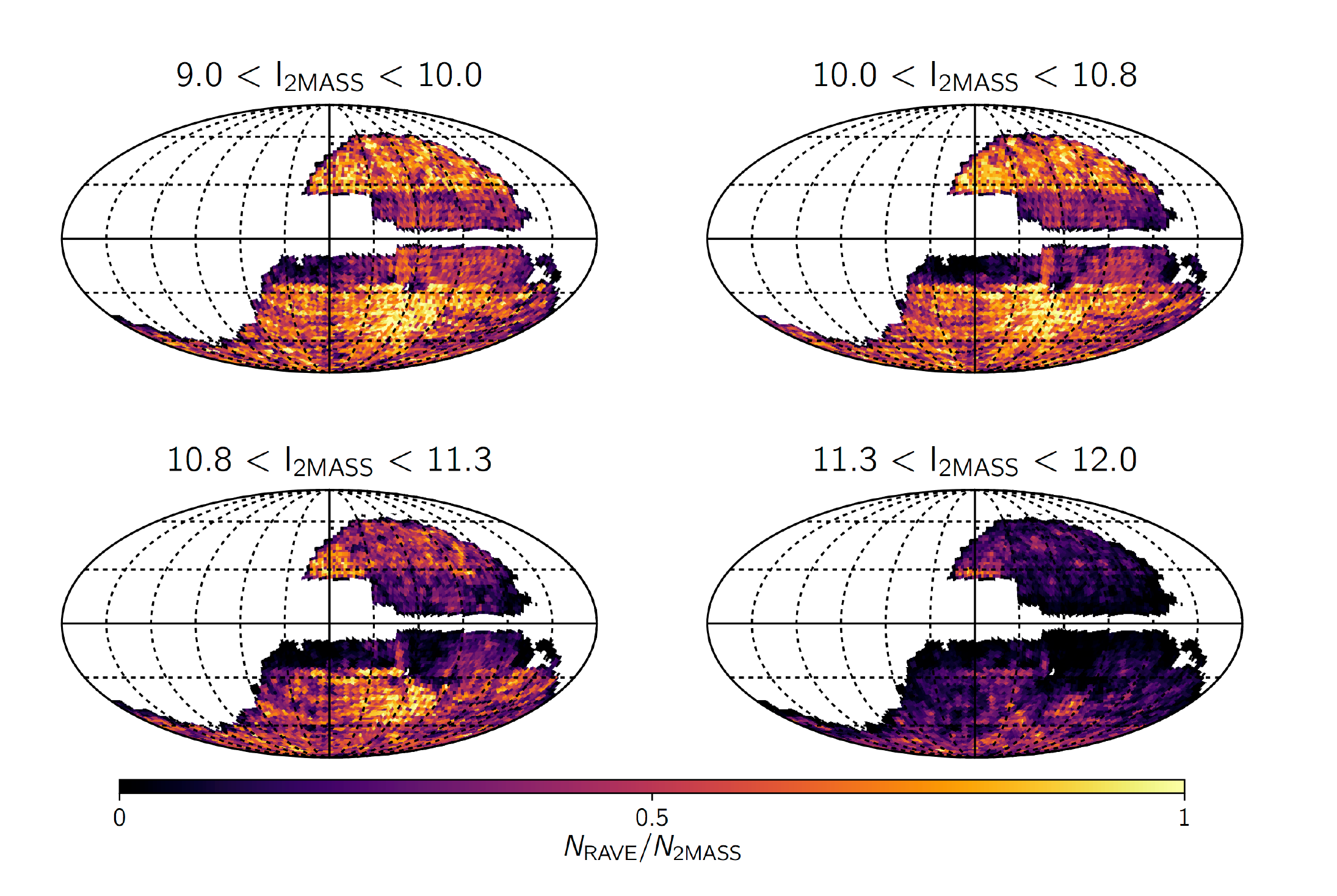} 
 \caption{Completeness (relative to 2MASS) of RAVE DR5 in Galactic coordinates as a function of observed magnitude bins . The \textsc{healpix} pixels are colour-coded by the fractional completeness, ($N_{\mathrm{RAVE}}/N_{\rm 2MASS}$). Adopted from Wojno \etal, 2007.}
   \label{fig1}
\end{center}
\end{figure}

\section{RAVE survey description}
The \textit{RAdial Velocity Experiment} RAVE\footnote{http://www.rave-survey.org} (Steinmetz \etal, 2006) is a magnitude-limited survey of stars randomly selected in the southern celestial hemisphere. The original design was to only observe stars in the interval $9<I<12$ but owing to the IR based selection function, some stars that are brighter and fainter can be found. The spectra are obtained from the 6dF facility on the 1.2m Australian Astronomical Observatory's UK-Schmidt telescope in Siding Spring, Australia, where three field plates with up to 150 robotically positioned fibres are used in turn. RAVE observations began in April 2003 and were completed in April 2013. Five data releases have been published. The most recent data release (DR5 Kunder \etal, 2017) presents radial velocities from 520,781 spectra of 457,588 unique stars, of which 255,922 stellar observations have parallaxes and proper motions from the Tycho-Gaia astrometric solution (TGAS) in Gaia DR1 (Gaia collaboration \etal 2016).

The astrometry and parallaxes from the first Gaia data combined with the RAVE DR5 radial velocities ensure that 10 km s$^{-1}$ uncertainties in space velocities for 70\% of the RAVE-TGAS stars can be derived.

The effective resolution of RAVE is $R=\lambda/\Delta \lambda \sim7500$ and the wavelength range coverage is around the infrared ionized Calcium triplet (IR Ca{\footnotesize II}, $\lambda \lambda 8410-8795$\AA), one of the widely used wavelength ranges for Galactic archeology. It is also the wavelength range in which the Gaia RVS is operating. Based on RAVE spectra, stellar parameters could be derived for 94\% of the sample, abundances of Al, Si, Ti, Fe, Mg and Ni are obtained for approximately 2/3 of the RAVE stars (see Table 1 for detailed numbers). These are generally good to $\approx 0.2$ dex, but their accuracy varies with SNR and, for some elements, depends also on the stellar population. Spectro-photometric distances are derived using a Bayesian pipeline (see McMillan \etal, 2017, and references therein).

RAVE was among the first spectroscopic surveys in Galactic astronomy with the explicit purpose of producing a homogeneous and well-defined data set. To achieve this goal, the initial target selection was based purely on the apparent $I$-band magnitudes of the stars. Indeed, a detailed analysis of the selection function of RAVE demonstrates that for stars brighter than I = 12, between $4000 \rm K < T_{\rm eff} < 8000 \rm K$ and $0.5 < \mathrm{log}\,g < 5.0$, RAVE is kinematically and chemically unbiased with respect to expectations from mock surveys such as \textsc{Galaxia}. The completeness of RAVE DR5 is shown in Fig.~1.

RAVE has meanwhile been complemented by a number of catalogs derived from RAVE spectra using alternative methods, for example the RAVE-on catalog featuring atmospheric parameters, element abundances, and ages using The Cannon (Casey \etal, 2017), a data driven approach trained on APOGEE (for giants) and K2/EPIC data (for sub giants and main sequence). Valentini \etal (2007) derived a recalibration for $\log g$ of RAVE giants in the color interval $0.50 < (J-K_S) < 0.85$ based on astroseismic $\log g$ determinations of 87 stars in the Campaign 1 of the K2 mission.
Jofre \etal (2017) applied the twin method to determine parallaxes to 232 545 stars of the RAVE survey using the parallaxes of Gaia DR1 as a reference. That way, twins could be found for 60 per cent of the RAVE sample that are not contained in TGAS or that have TGAS uncertainties that are larger than 20 per cent. Matijevi\u{c} \etal\ (2017) derived a catalogue of very metal poor stars on RAVE featuring 877 stars with at least a 50\% probability of being very metal-poor ([Fe/H] $< -2$ dex), out of which 43 are likely to be extremely metal-poor ([Fe/H] $< -3$ dex).
\begin{table}[t]
\begin{center}
\caption{Contents of RAVE DR5 } 
\label{ravegcs}
\begin{tabular}{ p{6.5cm}p{2.5cm}}
 & in DR5  \\ 
\hline
\hline
RAVE stellar spectra & 520,781 \\
Unique stars observed & 457,588 \\
Stars with  stellar parameters & 428,952 \\
Stars with elemental abundances & 339,750 \\  
\hline
TGAS + RAVE stellar spectra/unique stars & 255\,922 / 215,590 \\
\hline
\hline

\end{tabular}
\end{center}
\end{table}

\section{Galactic dynamics application}

First applications exploiting the combined astrometric/spectroscopic potential of RAVE + TGAS address the wave-like patterns seen in the streaming motions in the extended solar neighborhood with surveys like RAVE, LAMOST and SEGUE (see article by Widrow in this volume). Carrillo \etal\ (2017, see also article in this volume) demonstrated that the structure of the vertical velocity field is considerably affected by systematics of the respective proper motion catalogue. Using the TGAS values, a picture emerges that sees the vertical velocity field as a combination of bending and breathing modes, likely generated by  external as well as internal  mechanisms, i.e., a superposition of more than one wave existing simultaneously in the Milky Way disc. Wojno (2017, see also article in this volume) employed ages as derived from the distance pipeline with TGAS parallax priors by McMillan \etal (2017) to demonstrate that radial velocity gradients in the solar suburb depend on the age of the underlying stellar population with this gradient being noticeably
steeper for young metal-rich stars.

\begin{figure}[t]
\begin{center}
 \includegraphics[width=4in]{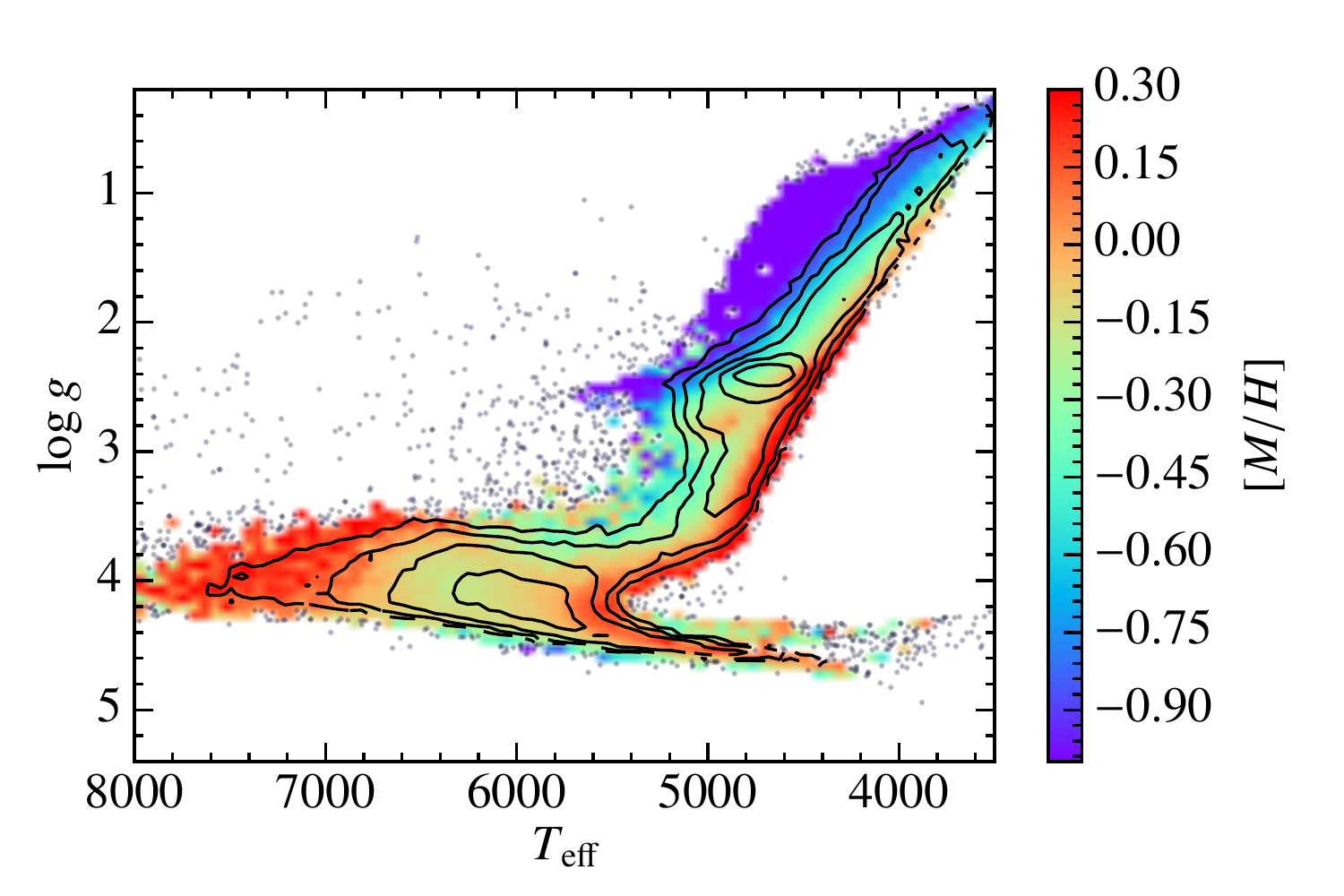} 
 \caption{$T_{\rm eff}$ and $\log g$ values of RAVE+TGAS stars derived using RAVE spectra and complementary photometry based on the Bayesian method with TGAS priors as described in McMillan \etal\ (2017). Pixels are colored by median metallicity, and overlaid contours show the density (with a logarithmic scaling in density between contours). Adopted from McMillan \etal, 2007.}
   \label{fig2}
\end{center}
\end{figure}

\section{Towards the 6th RAVE data release}

The spectro-photometric distance estimation based on the spectral data of stars in the fifth RAVE data release was considerably improved by combining them with parallaxes from the first Gaia data release. The combined distance estimates have been shown to be more accurate than either determination in isolation (McMillan \etal, 2017). With the considerably improved parallaxes of the second Gaia data release scheduled for April 2018, estimates for $T_{\rm eff}$ and $\log g$ can be taken from the Bayesian pipeline. The addition of parallax priors thus allows to partially lift the $T_{\rm eff}$-$\log g$ degeneracy inherent in the RAVE spectroscopic data and thus to derive considerably improved stellar parameters and, subsequently, abundances. This is demonstrated in Fig.~2., which shows the resulting $T_{\rm eff}$-$\log g$ diagram for the DR5+TGAS sample and distance errors roughly eight times larger than those anticipated for Gaia DR2. Clearly the red clump, blue horizontal branch and a binary sequence can clearly be identified. 

It is our plan to use this method in an iterative fashion with the RAVE spectroscopic pipeline to improve the accuracy of our stellar parameters and therefore also of the RAVE abundance estimates (see also contribution by G. Guiglion in this volume). Furthermore, full spectra data for all stars in RAVE DR6 will be provided.

\end{document}